# Efficient Determination of Equivalence for Encrypted Data

Jason N. Doctor, Jaideep Vaidya, Xiaoqian Jiang, Shuang Wang, Lisa M. Schilling, Toan Ong, Michael E. Matheny, Lucila Ohno-Machado, and Daniella Meeker

- J. Vaidya is with the Management Science & Information Systems Department, Rutgers University, Newark, NJ. Email: jsvaidya@business.rutgers.edu.

- X. Jiang, S. Wang, and L. Ohno-Machado are with UCSD Health Department of Biomedical Informatics, UC San Diego, La Jolla, CA. Email: x1jiang@ucsd.edu, shw070@ucsd.edu, lohnomachado@ucsd.edu.

- L. Schilling and T. Ong are with the Department of Medicine, University of Colorado, Anschutz Medical Campus, CO. Email: lisa.schilling@ucdenver.edu, toan.Ong@ucdenver.edu.

- M. Matheny is with the Geriatric Research Education and Clinical Care Service, Tennessee Valley Healthcare System VA, Nashville, TN, and the Department of Biomedical Informatics, Medicine, and Biostatistics, Vanderbilt University Medical Center, Nashville, TN. Email: michael.matheny@vanderbilt.edu.

- D. Meeker is with the Keck School of Medicine, University of Southern California, Los Angeles, CA. Email: dmeeker@usc.edu.

Schaeffer Center for Health Policy and Economics, University of Southern California, 635 Downey Way, Los Angeles, CA 90089-3333. T: 213.821.8142; F: 213.740.3460; jdoctor@usc.edu (corresponding author)

**Abstract**

Secure computation of equivalence has fundamental application in many different areas, including healthcare. We study this problem in the context of matching an individual's identity to link medical records across systems. We develop an efficient solution for equivalence based on existing work that can evaluate the "greater than" relation. We implement the approach and demonstrate its effectiveness on data, as well as demonstrate how it meets regulatory criteria for risk.

**Keywords**

Secure computation, encrypted data, set intersection, socialist millionaires' problem, record linkage

## 1. Introduction

Today, data are often collected across multiple systems and organizations. Much of these data are sensitive, and therefore need to be kept protected. Indeed, laws such as HIPAA and the EU General Data Protection Regulation (2016/679) require that data be safeguarded and only used in accordance to the purpose for which it was collected. Data are typically kept in silos, and only utilized in a limited fashion. Yet, the data can be of great value when linked together for many different applications. For example, if credit card transaction data across different banks are analyzed together, it may be possible to build better models for identifying fraudulent transactions. Similarly, in healthcare, if patient record data across the different institutions where the patient receives service may be aggregated, then it is possible to provide personalized care that takes into account all of the factors underlying the case. However, such analysis requires the integration and linkage of sensitive data across organizations. It may be useful to link records together when they belong/refer to the same patient, but clearly not otherwise. It is important to be able to identify equivalence between individual identifiers to determine whether the records refer to the same patient and can be merged into a single record. This has other applications as well, such as the case where two traders want to know if their buying and selling prices match in the absence of a trusted broker or the case where two agents want to authenticate that their shared secrets match to confirm their communications were not intercepted by an outside person or group. In all of these examples, identification of equivalence is needed so that private information is not revealed to a third party. Finding such equivalence in an efficient and secure manner is the question addressed in this work.

Private data are most often evaluated through sharing of information on private data systems. Such sharing occurs in social networks, online auctions, and between messaging and health information systems. Protection of sensitive data in these systems is difficult. Once data for comparison are transmitted to a private server, they are often evaluated in cleartext (i.e., unencrypted text) behind a firewall. However, an adversary may exploit vulnerabilities of the firewall to obtain the cleartext data behind it. More advanced approaches may involve a private server maintained by an honest broker who links data from different sites by a common hash ID with other data remaining in cleartext (Kho et al. 2015). High rates of re-identification are possible from even just a few disclosed values in "de-identified" data (de Montjoye et al. 2015).

Linking data from different sources increases the chance of re-identification to an even greater extent than single-source data storage because it reveals additional information. In contrast to private servers that rely on security secrets to act as safeguards, public security systems benefit from openness—adversaries have difficulty circumventing the security system because the system's workings are public knowledge. The system must be secure by design. There are no, or at least few, secrets to act as vulnerabilities.

Secure multiparty computation is a public approach to protect sensitive data. In this method, data are encrypted as private inputs to a public function that assigns an evaluation result such that only the result is known and the private inputs are not decrypted to cleartext during evaluation. This means comparisons are secure and limited by computation time and may be possibly constrained by communication. In essence, all work on encrypted data is carried out in public.

Yao's millionaires' problem, where two millionaires wish to determine which of them has more money without disclosing their fortunes, was the first example of the secure two-party method (Yao 1982). This work and many articles that followed address strict inequality (Cachin and Christian 1999; Ioannidis and Grama 2003; Shundong et al. 2008). Of note, Lin & Tzeng (2005) showed that intersection of two sets each generated by a special coding of private inputs could evaluate the "greater than" relation in a two-round highly efficient protocol (see Theorem 1). In Corollary 1, we show that repeated application of

Theorem 1 implies the equivalence relation, also called "socialist millionaires' problem"[1], but requires four rounds. Following this corollary, we prove that there is a simple two-round protocol that identifies a matching case under this special coding (Theorem 2). Application of the protocol in Theorem 2 entails far less computational time and fewer communication rounds than the four-round protocol (Corollary 1). Our approach also improves upon other greater-than-two round protocols for equivalence (e.g., Brandt and Felix (2006)).

In the approach we describe, we consider two parties – "Alice" and "Bob". Bob returns to Alice ciphertext of two consecutive integers: Bob's target integer, $n$, for comparison and one less than the target integer, $n - 1$. From this information, Alice can determine if the target integer matches her own integer in only two rounds. In this paper, we review the set intersection approach to secure multiparty computation. We then develop an efficient approach to computing equivalence in this context (Theorem 2). Finally, we illustrate the approach through record linkage of medical records across systems and evaluate its performance.

## 2. Problem Statement and Structural Assumptions

We first present the problem statement and then the notation and structural assumptions made in the rest of this work.

**Problem Statement**

*Two parties, Alice and Bob, have their private inputs x and y. Together, they would like to evaluate if the two inputs are equal, i.e., if x is the same as y, without revealing their inputs to each other or to any third party.*

Note that once equivalence can be computed for a pair of values, the same protocol can be used repeatedly to compute equivalences between sets of values. A non-trivial extension is to the case with multiple parties, where parties $P_1, \ldots, P_k$, may have values $v_1, \ldots, v_k$, and together wish to determine if $v_1 = \ldots = v_k$? We discuss this extension at the end of the article.

**Notation and Structural Assumptions**

*Representation of Numbers and Order Relations*

Throughout the paper we deal exclusively with integer-based private data (such as cell phone numbers). Our results extend easily though to any countable data such as fixed-point numbers (e.g., dollars and medical laboratory tests). Further, anything that has order is countable; therefore, our method also extends to any data that can be ordered (e.g. alphabetic text).

Integers identify private information as a 32-bit binary string. A 32-bit binary string is a group of *0* and *1* symbols that count positive integer positions. The string begins with "000…01", corresponding to the positive integer "1", and through binary counting after a digit reaches 1, an increment resets it to 0 but also causes an increment of the next digit to the left. For example, the 32-bit binary string for the integer "2" is "000…010", for "3" is "000…011" and so on.

We draw on the properties of order relations for proofs. The "greater than or equal to" relation, $\geq$, over private data is a total order: antisymmetric (if $x \geq y$ and $y \geq x$ then $x = y$), transitive (if $x \geq y$ and $y \geq z$ then $x \geq z$) and complete (for every $x$ and $y$, either $x \geq y$ or $y \geq x$ or both). Strict inequality $>$ follows if and only if $x \geq y$ and $x \neq y$.

---

[1] For the sake of social equality, two millionaires wish to determine if their fortunes are equal without disclosing their net worth (Boudot et al. 2001).

*Encryption Tables*

Alice's private input for a comparison is represented in an encryption table. This table is composed of 64 squares arranged in 32 vertical columns and two side-to-side rows. Columns are identified by positions 1 to 32 (inverted). The numbers 0 and 1 identify rows (because the table is a representation of a binary string for Alice's number). Each square has a numeric value, which is hidden from Bob through encryption. These numeric values encode Alice's number by their position in the table. Encryptions are homomorphic to a mathematical operation ○ (e.g., addition or multiplication). That is, evaluations of mathematical operations produce the same logical results when operating on encrypted data as they would be in cleartext implementations. Alice encrypts identity elements, $e$, in the row and column positions that correspond to her number (See Box 1); otherwise she places random numbers in the cells. These numbers are in the same range as those resulting from encryption.

**Box 1.** Encrypted table representing the number "1" as a 32-bit binary string (000...01) inverted to be read right to left, where $r$'s are random numbers, E() is an encryption function and $e$ is the identity element.

In Box 1, because the $r$'s are random and serial encryption of the same number assigns a new ciphertext each time the number is encrypted, each cell contains a unique ciphertext, provided the encryption system is semantically secure.

|   | 32 | 31 | 30 | 29 | 28 | 27 | 26 | … | 1 |
|---|----|----|----|----|----|----|----|----|----|
| **0** | r | E($e$) | E($e$) | E($e$) | E($e$) | E($e$) | E($e$) | … | E($e$) |
| **1** | E($e$) | r | r | r | r | r | r | … | r |

The approaches that we will discuss each rely on set intersection between sets of substrings of the binary string. Note that if Bob multiplies encrypted cell entries as a means of identifying his substrings and sends these multiplications as messages to Alice for decryption, the products will be homomorphic to ○ upon decryption. Further, since $e \circ e = e$, if one or more decryptions is $e$, then Bob has a substring corresponding to that of Alice. Next we formalize this procedure and extend it to the equality relation.

We assume a Paillier cryptosystem (Paillier and Pascal, 1999), although any partial or fully homomorphic public/private key cryptosystems can be used. The Paillier cryptosystem has the following useful homomorphic property discussed earlier:

PROPERTY 1: Multiplying encrypted messages results in the addition of the original plaintexts mod $n$. Formally: $DE(m_1) \cdot E(m_2) \mod n^2) = m1 + m2 \mod n$, where $m_1$ and $m_2$ are messages, $n$ is a natural number, E() is the Paillier encryption algorithm and D() is the decryption algorithm.

We will exploit Property 1 to identify matching messages. The definitions and theorems that are part of our main results are necessary for this task.

**3. Main Result**

We first review Lin & Tzeng's (2005) result for secure computation of strict inequality (Theorem 1). We then show how it implies equivalence under repeated evaluation (Corollary 1), but is inefficient requiring 4 rounds. Finally, we develop a 2-round protocol for equality using this special coding in two rounds (Theorem 2). The following two definitions are a necessary prerequisite to the theorems.

DEFINITION 1. A *one-encoded set* of a binary string contains all sequences beginning at string position, i, that starts with 1 and end with the last position *n* (far left) of the string.

For example, the one encoded set for the number 3, or binary string "110," is {1, 11}.

DEFINITION 2. A *zero-encoded set* of a binary string contains all sequences beginning at string position, j, that starts with 0, replacing that position with 1 and end with the last position *n* (far left). We ignore strings for positions that start with 1.

For example, the zero-encoded set for the number 2, or binary string "010" is {1, 011}.

THEOREM 1 (Lin & Tzeng, 2005). Let $x$ and $y$ be positive integers. $x > y$ *if and only if* the one-encoded set for $x$ and the zero-encoded set for $y$ share a common set element.

*Proof:* A proof is given in Lin & Tzeng (2005).

COROLLARY 1. Let $x$ and $y$ be positive integers; if each of the intersections of the one-encoded set for $x$ with the zero-encoded set for $y$ and the one-encoded set for $y$ with the zero-encoded set for $x$ are the null set, then $x = y$.

*Proof*: By Theorem 1, if the intersection of the one-encoded set for $x$ and the zero-encoded set for $y$ is the null set, then $\neg(x > y)$ and either $x = y$ or $y > x$. Also, by the same Theorem, if the one-encoded set for $y$ and the zero-encoded set for $x$ is the null set, then $\neg(y > x)$ and either $x = y$ or $x > y$. Since, $x$ cannot be both strictly greater and strictly less than $y$, it follows that $x = y$. Q.E.D.

It is clear from Corollary 1, that equivalence under Lin & Tzeng (2005) entails four rounds and additional secure computing structure: 1) Alice computes the one-encoded set for $x$, 2) Bob returns the zero-encoded intersection with $y$, 3) Alice computes the zero-encoded set for $x$, and 4) Bob returns the one-encoded intersection with $y$. After four rounds, Alice knows if $x > y$ and if $y > x$. And, $\neg(x > y)$ and $\neg(y > x)$ implies $x = y$ (Corollary 1). Although this use of Theorem 1 to evaluate equivalence is possible, it is inefficient, requiring Alice to encrypt her $x$ in two different ways, which is computationally intensive.

Building on these results from Lin and Tzeng (2005), we show that a more efficient protocol is possible. This protocol requires only two rounds of computation to identify equivalence, while preserving the advantages in efficiency of Lin and Tzeng's technique.

THEOREM 2. Let $x$ and $y$ be positive integers. The following two statements are equivalent:

(i) $x = y$.

(ii) The one-encoded set for $x$ shares a common element with the zero-encoded set for $y - 1$, but not with the zero-encoded set for $y$.

*Proof:* We prove that (i) $\Rightarrow$ (ii) and (ii) $\Rightarrow$ (i). To show (i) implies (ii), we note that if $x = y$, then $x > y - 1$ and the one-encoded set for $x$ and the zero-encoded set for $y - 1$ share a common set element (Theorem 1). To see that the one-encoded set for $x$ does not intersect with the zero-encoded set for $y$ we observe that by hypothesis of the contrapositive of Theorem 1, $y \geq x$ if and only if the one-encoded set for $x$ and the zero-encoded set for $y$ do not have a common element, which by definition makes their intersection the null set. This establishes that (i) $\Rightarrow$ (ii). To show that (ii) implies (i), observe that (ii) implies that $x > y - 1$ (Theorem

1) from which the weaker statement $x \geq y - 1$ also holds. Since $x - 1$ is the largest positive integer less than $x$ we can write: $x > x - 1 \geq y - 1$. From $x - 1 \geq y - 1$, it follows that $x \geq y$. We are left to show that $y \geq x$. Suppose now that $x$ is not equal to $y$ nor is it the largest positive integer less than or equal to $y$, then $y \geq y - 1 > x$, or $y - 1 > x$ which is absurd because by (ii) and Theorem 1, $x > y - 1$. Therefore, $y \geq x$. Since, $y \geq x$ and $x \geq y$, $x = y$. Hence, (ii) ⇒ (i) and (i) ⇒ (ii). Q.E.D.

The secure two-party protocol for implementing Theorem 1 also applies to Theorem 2. It is described in detail in Section 3 of Lin & Tzeng (2005). Here we briefly review the protocol. First, Alice generates an encryption table for her value (See Box 1) and she sends this to Bob. Second, Bob zero-encodes his number producing a set of substrings. Third, using Alice's table, Bob identifies the encryptions and/or random numbers that correspond to his substrings and multiplies them together, sending the products back to Alice as messages. He sends two sets of messages one for his target number, $n$, and another for $n - 1$ along with random "filler" messages to prevent information leakage. Fourth, Alice decrypts these messages. Under additive homomorphic encryption, if at least one message for a number returns "0" upon decryption and no decrypted messages return "0" for the other number, then Alice and Bob's number are equivalent. The detailed protocol is presented in Algorithm 1.

---

Algorithm 1 (Equivalence)

---

1: Inputs: Alice with a number $x$ and Bob with a number $y$, where both x and y are in the range [0 n]
2: Output: True if and only if $x = y$

3: Alice creates an encryption table W of size (log n x 2), based on the binary representation $b$ of her value $x$ as follows:
4: if the $i^{th}$ bit $b_i = 1$, then
5:     set $W_{i1} = E(0)$, a random encryption of 0
6:     set $W_{i0} = r$, a randomly chosen value from the range of the encryption E
7: else
8:     set $W_{i0} = E(0)$, a random encryption of 0
9:     set $W_{i1} = r$, a randomly chosen value from the range of the encryption E
10: endif
11: Alice sends W to Bob {Note that W is an alternative form of the 1-encoding of $x$ obscured to Bob}
12: Bob creates the 0-encoding of $y$ and $y$-1, denoted as sets $y_0$ and $y'_0$
13: Bob: For each string in $y_0$, multiply the corresponding values of W, permute and send to Alice
14: Bob: For each string in $y'_0$, multiply the corresponding values of W, permute, and send to Alice
15: Alice: Decrypt the set of values received.
16: Alice: If there is at least one 0 in one of the sets of values but not in the other, output true to Bob, otherwise output false to Bob

---

*Extension to more than two parties*

---

The extension of Alice and Bob 2-party secure computation to identify equivalence among $n$-parties, for $n > 2$, is a linear first order recurrence relation: increasing comparisons from $a_m$ for $m$ pairs to $a_{m+1}$ for $m + 1$ pairs of sites, requires $a_{m+1} = a_m + 2$ comparisons. By adding a pair of sites we increase the number of two-party comparisons by 2 (i.e., in serial order, each new party compares its data with Alice who has made a subset of her data based on comparisons with previous parties, or, the new pair compares its data together, subsets it and compares to the subset of Alice's data with Bob). Obviously, if we just add a single site and not a pair we increase the number of two-party comparisons by 1. Alice may orchestrate comparisons among sites without the need for an honest broker.

## 4. An Illustrative Example

For simplicity, we give an example using an abbreviated 3-bit binary string, Theorem 2 together with the Paillier cryptosystem and an encryption table to identify matches: Alice and Bob each hold a case with the unique identifier "5" and would like to identify if this is a matching record.

For each binary position in the 2 x 3 table Alice uses Paillier cryptosystem to encrypt the binary position which matches that of her string with the additive identity element, 0, E(0), otherwise she inserts a random number, $r$. The $r$'s are of no concern to Alice. She does not need to know what they are. For example, for the number "5":

**Table 2.** Encrypted 2 x 3 table for the number "5".

5    Binary string 101

|   | 3 | 2 | 1 |
|---|---|---|---|
| **0** | $r$ | E(0) | $r$ |
| **1** | E(0) | $r$ | E(0) |

Alice sends this table to Bob. Bob conducts a mathematical operation on Table 2's ciphertext following the steps of the protocol with respect to his unique identifier (also "5") and one less than his unique identifier 5 - 1 = 4. That is, using Theorem 2, Bob will send Alice messages on "4" and "5" so that she can determine if there is a match.

Specifically, Bob considers his zero-encoded sets for the numbers, "4" and "5". The zero-encoded set for "4" is {11, 101} and the zero-encoded set for "5" is {11}. Bob then uses Table 2 to generate encrypted messages by selecting table cells that correspond to his zero-encoded sets for his numbers and multiplying them. He sends these sets to Alice, separating messages by number.

For example, let $m$ be a message. For the number $y - 1 = 4$, Bob computes $m_1 = E(0) \cdot E(0) \cdot E(0)$ for "101" and he computes $m_2 = E(0) \cdot r$ for "11". For the number $y = 5$, he computes $m'_1 = E(0) \cdot r$ for "11".

Bob then sends Alice $m_1$, $m_2$ and $m'_1$; coding these messages so that $m_1$ and $m_2$ are considered together and $m'_1$ is considered separately.

Alice then decrypts these messages. By Property 1 of Paillier cryptosystem, she learns that $m_1 = D(E(0) \cdot E(0) \cdot E(0)) = 0 + 0 + 0 = 0$, and $m_2, = D(E(0) \cdot r) = 0 + D(r) = s > 0$. She concludes, based on Theorem 1, that there is a common element between her $x$ and one of Bob's numbers, and that her $x$ is greater than one of the numbers Bob sent since $m_1 = 0$ (a perfect substring match). She then decrypts $m_1$' and concludes that her $x$ and Bob's other number are different. By Theorem 2 and the fact that one consecutive number is strictly greater and the other is not, she deduces that Bob's target number and her number are equivalent.

## 5. Implementation Details and Experimental Evaluation

We implement the algorithm described above with respect to Theorem 2. However, to reduce the number of comparisons we implement a modified Merge Sort algorithm (Goldreich, Micali & Wigderson (1987);

McCool et al. (2012)) that reduces *a* x *b* paired comparisons matrix to *a* + *b* comparisons to sort the data. The algorithm proceeds as follows:

(i) Alice and Bob each sort their data from least to greatest in separate arrays noting ties.

(ii) They compare least previously not compared values.

(iii) If Alice's least previously not compared value is greater than Bob's, then his array position increments one for the comparison.

(iv) If Bob's least value is greater (which can be deduced by the results), then Alice's value increments.

(v) If Alice and Bob have equal values, each site advances to the next value that is greater than the one that produced the equality.

Since each local site orders its data from least to greatest, for each comparison the algorithm is able to place an observation. We note that this process could leak information if carried out without obscuring the position of Alice's number in the request or Bob's result. To prevent Alice from leaking information, the following approach may be used: 1) Alice and Bob order their data least to greatest; 2) Alice *randomly* selects a number from her vector to compare with Bob's lowest number; 3) Alice then uses a binary search to home in on a match along her vector. This removes any lawfulness in Alice's choice of her next number.

Bob may only leak 'greater than' information. We discuss in the section labeled "Security Analysis" a simple way to eliminate this problem is by having Alice and Bob each use the same cryptographically secure (keyed) hash function to compute the hashes of each of their values before carrying out the protocol. We note here, however, that one other way to prevent Bob from exposing any information beyond the match result, is for him to exponentiate each of his encrypted multiplications to base zero subtracting the (raised to the base zero) encrypted product resulting from comparison of Alice's number with his *y* from the (raised to the base zero) result of the comparison with his *y* - 1.[2] The codomain of this function is either 1 ('match') or 0 ('no match'). To carry out this calculation on encrypted data requires fully homomorphic encryption (Gentry (2009)) and bit decomposition of the encrypted products (Brakerski, Gentry & Vaikuntanathan (2014)). We do not pursue empirical analysis of this fully homomorphic approach here in favor of describing a simpler and efficient hashing solution.

*Data Set*

The synthetic datasets used in this paper were generated using the Mockaroo synthetic data generator (www.mockaroo.com). The data fields included in both datasets are First name, Last name, Social security number (SSN), Date of birth and Identification (ID). Each dataset contains 10K records and both datasets have 6K records in common. A record which is in both datasets has the same value in the ID field. Therefore, the values of the ID field are used as the true answer to verify linkage.

*Computing Environment*

The algorithm was carried out on an Apple, inc. MacIntosh computer with a 2.4 Ghz Intel Core i5 processor and 8 GB (1699 MHz DDR3) of memory. Software was written in the statistical computing language R, Version 3.3.0, released May 3rd, 2016 (Ihaka, Ross, and Robert 1996) and utilized the package HomomorpheR for additive homomorphic encryption (Narashimhan, 2016).

---

[2] Specifically, if *m* and *m'* are encrypted messages, then $f(m, m') = g(m') - g(m)$, where $g(z) = 0^z$.

*Results*

For Alice to produce encryption tables for 10,000 of her observations it took 63 minutes and required 462.5 MB of storage for a single vector of encrypted numeric identifiers. Dynamic multiplication of Alice's encryptions by Bob to represent his data as messages, decryption by Alice and determination of a match through sorting took 6 hours, 32 minutes and 27 seconds.

Computation time for each component on 100 representative observations is given in Table 3.

**Table 3**. Computation time (*hh*:*mm*:*ss*) by analysis component per 10,000 representative observations.

| Time | Alice encryptions | Bob's Message Generation | Decryption by Alice |
| --- | --- | --- | --- |
| User | *00:06:40* | *00:16:35* | *00:08:20* |
| System | *00:03:20* | *00:21:40* | *00:03:20* |
| Elapsed | *00:08:20* | *18:46:40* | *00:08:20* |

## 6. Protocol Analysis

We first analyze the computation and communication cost of the proposed protocol and then analyze its security.

*Computation and Communication Analysis*

The proposed protocol is quite efficient. First, note that for a single comparison, it requires only 2 rounds of interaction for Alice to learn the result, and three for Bob to do the same. In terms of the number of bits, Alice sends 2*log n* values, where each value comes from the range of the Paillier encryption used. Since there can at most log n 0-encoded bit strings, Bob can also send up to 2*log n values, where each value comes from the range of the Paillier encryption used. In terms of computation cost, Alice has to create at most log n random encryptions of 0 (these could be pre-computed), and perform at most 2*log n* Paillier decryptions. Bob has to carry out at most 2(*log* n)$^2$ Paillier multiplications. To get the final output over the entire set of records, we have at most $a + b$ comparisons carried out, thus providing a multiplicative factor of ($a + b$), which is linear to size of the data.

*Security Analysis*

We can prove that the proposed comparison protocol is secure under the framework of secure multiparty computation. We do make the assumption that both parties (Alice and Bob) are "semi-honest" agents. This means that these sites follow the protocol exactly, but can try to infer additional information from intermediate results. The proof of security follows a simulation based approach. i.e., we aim to show that the messages received by either Alice or Bob can be simulated by them while having access to only their inputs and the final result.

Note that the basic comparison protocol (Algorithm 1), as presented above, leaks more information than pure equivalence. Specifically, if *x* and *y* are not equal, the protocol tells us whether *x > y* or *y > x*. This can be seen from the following: the protocol essentially checks if *x > y* as well as if *x > y - 1*. If *x* is indeed equal to *y*, then precisely one of the above comparisons (*x* compared to *y* - 1) will return true. In this case, this process of double comparisons does not leak any additional information since this is precisely what we R. Ihaka, I. Ross, and G. Robert, "R: A Language for Data Analysis and Graphics," *J. Comput. Graph. Stat.*, vol. 5, no. 3, p. 299, 1996.
expect. However, if *x* compared to y also returns true, we now know that *x* is greater than *y*. Similarly, if *x* is not greater than *y* and *x* is not greater than *y* - 1, then it is clear that *y* is greater than *x*. However, as discussed above, once the data have been hashed with a cryptographically secure (keyed) hash function, determining order relationships among the hashes does not give any information regarding the actual data. We also note that Alice would prefer to use the protocol (with hashed data) over sharing hashed data directly with Bob because they could easily be decrypted with a shared key. Direct hashed data comparisons could only be resolved by introducing an honest broker, which is inefficient, costly in human capital and requires a high level of trust in a third party.

Apart from this leakage, which can be easily dealt with by hashing the data, the protocol leaks no information. Bob only receives a message from Alice at line 11, where all of the values come from the range of a semantically secure encryption algorithm. Since the encryption is semantically secure, this does not reveal any information to Bob. Bob, also sends the multiplied messages (in lines 13 and 14), which can be used to infer the last bit of the value, which however is a hash of the actual value, and therefore does not reveal anything regarding the data. Note that it is also possible for Bob to reverse the order of lines 13 and 14 without impacting the correctness of the protocol, since Alice only recognizes equivalence if precisely one of the messages in either case contains 0. However, if Bob randomly chooses the order, it is not now possible for Alice to know whether the true last bit was 0 or 1, thus protecting his data, even if the hash were not present.

**7. Conclusion**

We develop an approach to equivalence that may be useful for several applications. In addition to secure matching of buying and selling prices as well as authentication, the procedure may be used for identity matching in large data medical data sets. Here there is the advantage that Alice's encrypted computations may be done "off-line". In other words, Alice may build and store her encrypted data at any time prior to the matching exercise. In addition, the approach offers some efficiency advantages over direct application of the set intersection method described in prior work. Future studies should evaluate the feasibility of this analysis in large data systems.

**Acknowledgements**: We thank Kayleigh Barnes for assistance in identifying the zero information leak function presented in Footnote 2. The authors would like to also acknowledge pSCANNER, which is supported by the Patient-Centered Outcomes Research Institute (PCORI), Contract CDRN-1306-04819. Contract-PI: Lucila Ohno-Machado.